\documentclass[twocolumn]{jpsj2} 
\usepackage{bm}
\usepackage{graphicx,color}%
\usepackage{amsmath,amsfonts,amsthm,amssymb,slashbox}

\title{
\textit{Ab-initio} Low-Energy Model of Transition-Metal-Oxide Heterostructure LaAlO$_3$/SrTiO$_3$
}

\author{Motoaki \textsc{Hirayama}\thanks{E-mail: hirayama@solis.t.u-tokyo.ac.jp},$^{1,3}$ Takashi \textsc{Miyake},$^{2,3}$ and Masatoshi \textsc{Imada}$^{1,3}$}

\inst{$^1$Department of Applied Physics, University of Tokyo, 7-3-1 Hongo, Bunkyo-ku, Tokyo 113-8656, Japan\\
$^2$Nanosystem Research Institute, AIST, Tsukuba 305-8568, Japan\\
$^3$CREST, JST, 7-3-1 Hongo, Bunkyo-ku, Tokyo 113-8656, Japan}

\abst{
We develop the multi-scale \textit{ab-initio} scheme for correlated electrons (MACE) for transition-metal-oxide heterostructures, and determine the parameters of the low-energy effective model.
By separating Ti $t_{2g}$ bands near the Fermi level from the global Kohn-Sham (KS) bands of LaAlO$_3$/SrTiO$_3$ which are highly entangled with each other, we are able to calculate the parameters of the low-energy effective model of the interface with the help of constrained random phase approximation (cRPA).
The on-site energies of the Ti $t_{2g}$ orbitals in the $1$st-layer is about $650$ meV lower than those in the $2$nd-layer.
In the $1$st-layer, the transfer integral of the Ti $t_{2g}$ orbital is nearly the same as that of the bulk SrTiO$_3$, while the effective screened Coulomb interaction becomes about $10$ percent larger than that of the bulk SrTiO$_3$.
The differences of the parameters from the bulk SrTiO$_3$ reduce rapidly with increasing distance from the interface.
Our present versatile method makes it possible to derive effective \textit{ab-initio} low-energy models and allows studying interfaces of strongly correlated electron systems from first principles.
}

\kword{first-principles calculation, effective Hamiltonian, downfolding, constrained RPA method, correlated-electron systems, heterostructure, interface, two-dimensional electron systems}

\begin{document}
\maketitle


\section{Introduction}

In recent years, interfaces of strongly correlated electrons have been under intense investigations.
Especially, transition-metal-oxide heterostructure SrTiO$_3$ (STO)/LaAlO$_3$ (LAO) has received a lot of attention, because of its remarkable transport properties\cite{ohtomo}.
The interface of the SrTiO$_3$/LaAlO$_3$ shows metallic conductivity, although the bulk materials of each transition-metal-oxide, SrTiO$_3$ and LaAlO$_3$, are band insulators.
The TiO$_2$-terminated ($n$-type) interfaces show the metallic conductivity, when the thickness of deposited LAO layers is thicker than four unit cells \cite{thiel}, but the SrO-terminated ($p$-type) interfaces are insulating for any LAO thickness\cite{nakagawa}.  
The origin of this conductivity at the $n$-type interface remains a debated issue. 
An intrinsic electronic effect, namely the polar discontinuity, and an extrinsic atomic effect, atomic vacancies, were proposed to play key roles at high carrier concentrations.
When the polar LAO layers are deposited on the TiO$_2$-terminated substrate STO, an electric potential along the [$001$] direction diverges as the LAO thickness increases.
The instability of this electric potential is suppressed, if $1/2$ doped electron or a corresponding atomic charge vacancy per unit cell exist at the interface\cite{nakagawa,hirayama}.
Besides such transport properties, the existence of superconductivity and magnetic order are also reported at the SrTiO$_3$/LaAlO$_3$ interfaces\cite{reyren,brinkman}.
These unique transport properties are expected to offer a useful functionality in the possible applications.

The Local Density Approximation (LDA) calculations for transition-metal-oxide heterostructures have already been performed by several groups.
Park \textit{et al.} investigated LaAlO$_3$/SrTiO$_3$ for three kinds of superlattice structures with $n$-type, $p$-type, and both types of interfaces\cite{park}.
By taking into account the relaxation of lattice, Ishibashi and Terakura investigated the influence of LaAlO$_3$ thickness on the carrier density at the interface\cite{ishibashi}.
The LDA, however, often fails to capture correlation effects in transition-metal-oxides.
To study novel electronic phases in transition-metal-oxide heterostructures, in addition to the carrier doping from the electronic reconstruction, we should treat the correlation effects beyond the LDA, because the correlation effect is, on general grounds, expected to be enhanced at interfaces because of the effective reduction of the spatial dimensionality.

The LDA+U is useful for the strongly correlated materials, especially for insulators.
Pentcheva and Pickett tried to explain the insulating behavior of the $p$-type LAO/STO interface by introducing the on-site $U$, since the LDA gives a metallic state\cite{pentcheva}. 
However, in the LDA+U method, 
there is no established \textit{ab-initio} way to estimate $U$,  
although the value of $U$ strongly affects the calculated results of the low-energy physics.
Furthermore, owing to the experimental difficulties, $U$ has to be cited from the experimental values of the bulk, not of the interface, especially in the case where thick LAO layers are on a STO substrate.
Furthermore, the single-particle theory often collapses in strongly correlated materials, even with the suitable $U$.

To overcome such serious problems, a combined method of the multi-scale \textit{ab-initio} scheme for correlated electrons (MACE) is very useful\cite{imada}.
With the constrained random phase approximation (cRPA), we can calculate not only the transfer integral but also the screened Coulomb interaction without relying on any experimental parameters.
Correlation effects can be treated accurately by solving the obtained \textit{ab-initio} effective model using low-energy solvers such as variational Monte Carlo\cite{tahara}, dynamical mean field theory (DMFT)\cite{metzner, georges} and path-integral renormalization group (PIRG)\cite{imada2, mizusaki}.
The MACE has already been applied to a wide variety of materials; semiconductor\cite{nakamura1}, transition-metal\cite{solovyev1, solovyev2, miyake }, transition-metal-oxide\cite{miyake, pavarini, imai1, imai2, otsuka}, molecular organic conductors\cite{nakamura4} and Fe-based layered superconductor\cite{nakamura2, miyake2}. 
So far, the MACE has been applied to the bulk material.
Since interfaces have opened a new avenue of research, it is highly desired to develop a methodology which is able to treat the interface from the same footing of the MACE to understand the electron correlation effect from the first principles. 

In this paper, we apply the \textit{ab-initio} downfolding method to transition-metal-oxide heterostructures, and determine an \textit{ab-initio} low-energy effective Hamiltonian of the SrTiO$_3$/LaAlO$_3$. 
We derive the effective model from the conduction bands near the Fermi level originated mainly from the Ti $t_{2g}$ orbital.
The on-site energies of the Ti $t_{2g}$ orbitals in the $1$st-layer is about $650$ meV lower than those in the $2$nd-layer.
In the $1$st-layer, the transfer integral is nearly the same as that of the bulk SrTiO$_3$, while the effective screened Coulomb interaction becomes about $10$ percent larger than that of the bulk SrTiO$_3$.
The parameters of the $2$nd-layer are similar to those of the bulk SrTiO$_3$.

In Sec.2 we describe our method.
Section 3 describes the band structure and the derived effective model of the SrTiO$_3$/LaAlO$_3$. 
We also present an effective model of SrTiO$_3$ for comparison.
Section 4 is devoted to summary.


\section{Methods}

To construct the effective low-energy model of the heterostructure with well defined parameters, we treat the screening by the high-energy parts working on the low-energy parts starting from the band structure of LAO/STO, and renormalize the high-energy parts into the low-energy parts.
This downfolding procedure is introduced by Aryasetiawan \textit{et al.}\cite{aryasetiawan1} and Solovyev \textit{et al.} \cite{solovyev1}.
The low-energy effective model obtained from the downfolding procedure offers a starting point for studies on low-energy physics.
 
We consider an extended Hubbard Hamiltonian describing low-energy electronic properties of the interface of LAO/STO,
\begin{multline}
\mathcal{H} = \sum_{\sigma} \sum_{ij} \sum_{nm} t_{mn} (\bm{R_i}-\bm{R_j}) a_{in}^{\sigma \dagger} a_{jm}^{\sigma} \\
+ \frac{1}{2} \sum_{\sigma \rho} \sum_{ij} \sum_{nm} \biggl\{ U_{mn}(\bm{R_i}-\bm{R_j}) a_{in}^{\sigma \dagger}  a_{jm}^{\rho \dagger} a_{jm}^{\rho} a_{in}^{\sigma} \\ 
+J_{mn}(\bm{R_i}-\bm{R_j}) \bigl( a_{in}^{\sigma \dagger} a_{jm}^{\rho \dagger} a_{in}^{\rho} a_{jm}^{\sigma} +a_{in}^{\sigma \dagger} a_{in}^{\rho \dagger} a_{jm}^{\rho} a_{jm}^{\sigma}\bigr) \biggr\}, 
\label{eqH}  
\end{multline} 
where $a_{in}^{\sigma \dagger}$ ($a_{in}^{\sigma}$) is a creation (annihilation) operator of an electron with spin $\sigma$ in the $n$-th orbital, which is defined by a maximally localized Wannier function centered at the $\bm{R_{i}}$-th unit cell\cite{souza,marzari}.  
Especially, in this paper, we construct a model for all Ti $t_{2g}$ orbitals, not only at the interface but also in the bulk STO region, in the supercell of LAO/STO.
With thick LAO and STO layers, this $3$-dimensional model corresponds to the semi-infinite $3$-dimensional model for (vacuum)-(LaAlO$_3$)$_{N}$-(SrTiO$_3$)$_{\infty}$ where the LAO layers are deposited on the substrate bulk STO, because the inter-supercell screening effect is rather weaker than the intra-one.
Therefore, the parameters in this calculation may be used not only for superlattices but also for a semi-infinite interface LAO/STO.       
The parameters of the Ti $t_{2g}$ orbitals away from the interface recover the bulk STO nature in both the supercell and semi-infinite interface, because the polarizations away from the bulk STO region contribute little to the screening to such $t_{2g}$ orbitals.
In fact, in the present calculation, we show that the parameters in the $2$nd STO layer from the interface nearly converge to those of the bulk STO.
With thick LAO layers, the parameters of this $3$-dimensional model are also similar to those of a $2$-dimensional model for LAO/STO superlattice, where all of the inter-supercell screening effects are renormalized.

We derive the effective model in the following way:
First, we calculate the whole band structure of the transition-metal-oxide heterostructure in the framework of density functional theory (DFT).
We then choose the target low-energy band around the Fermi level and construct the maximally-localized Wannier functions\cite{souza} in the low-energy Hilbert space.    
The transfer integral, which defines the one-body part of the low-energy model, is obtained as the matrix element of the Kohn-Sham (KS) Hamiltonian ${\cal{H_{{\rm KS}}}}$,   
\begin{equation} 
t_{mn}(\bm{R}) = \langle \phi _{\bm{0}m} | {\cal{H_{{\rm KS}}}} | \phi _{\bm{R}n}\rangle , 
\end{equation} 
where $\phi _{\bm{R}n}$ is the $n$-th Wannier function centered at the cell $\bm{R}$.

Next, we renormalize the effect from the high-energy space ($r$ space) into the target low-energy space ($d$ space), and evaluate effective electron interaction by using the cRPA.
We divide the polarization
\begin{multline}
P(\bm{r},\bm{r'};\omega )=\sum _{i} ^{occ} \sum _{j} ^{unocc} \psi _i(\bm{r})\psi _i ^{\ast}(\bm{r'})\psi _j ^{\ast}(\bm{r})\psi _j (\bm{r'}) \\
      \times \Bigl[\frac{1}{\omega -\epsilon _j +\epsilon _i +i\delta }-\frac{1}{\omega +\epsilon _j -\epsilon _i -i\delta }\Bigr]
\label{polarization}
\end{multline}
into the polarization $P_d$ that includes only the $d$-$d$ transitions and the rest of the polarization $P_r$.
The screened Coulomb interaction $W$ is given by
\begin{equation}
\begin{split}
W&=[1-vP]^{-1}v \\
 &=[1-W_rP_d]^{-1}W_r,  
\label{screenedW}
\end{split}
\end{equation}
where we define the partially screened Coulomb interaction $W_r$ that does not include the screening arising from the polarization from the $d$-$d$ transitions as 
\begin{equation}
W_r=[1-vP_r]^{-1}v.
\label{Wr}
\end{equation}
The screened Coulomb matrix is defined by
\begin{multline}
W_r(\bm{R_1}n,\bm{R_2}n',\bm{R_3}m,\bm{R_4}m';\omega ) \\
 = \int d^3rd^3r'\phi _{\bm{R_1}n} ^{\dagger }(\bm{r})\phi _{\bm{R_2}n'}(\bm{r}) \\
     \times W_r(\bm{r},\bm{r'};\omega )\phi _{\bm{R_3}m} ^{\dagger }(\bm{r'})\phi _{\bm{R_4}m'}(\bm{r'}).
\label{Wrrep}
\end{multline}
The effective Coulomb interaction $U$ and the exchange interaction $J$ are given by
\begin{equation}
U_{nm}(\bm{R}) = W_r(\bm{0}n,\bm{0}n,\bm{R}m,\bm{R}m;0 ),
\end{equation} 
\begin{equation}
J_{nm}(\bm{R}) = W_r(\bm{0}n,\bm{0}m,\bm{R}m,\bm{R}n;0 ).
\end{equation}

In LaAlO$_3$/SrTiO$_3$, the Ti $t_{2g}$ states are entangled with the La $4f$ states. 
We disentangle the Ti $t_{2g}$ bands using the recently developed disentangling technique \cite{miyake3}.

In the LDA, the La $4f$ level in LAO/STO is located near the Fermi level.
However, in the real material, the La $4f$ bands are expected to be at higher energy, and, as a consequence, do not strongly screen the interactions between the electrons in the $t_{2g}$ bands.
To calculate the parameters of the low-energy model with a higher accuracy, a better way is to take into account the correlation effect of the La $4f$ bands by the GW approximation (GWA) as a preconditioning before determining the effective model for Ti $3d$ $t_{2g}$ orbitals.
The detail of the GWA for the disentangled $4f$ bands is explained in Appendix. 

Computational conditions are as follows.
We calculate the band structures of the bulk STO and LAO, and the LAO/STO heterostructure based on the DFT-LDA 
 \cite{hohenberg, korn}.
The calculations are carried out with the program based on the full-potential linear muffin-tin orbitals (FP-LMTO) method \cite{methfessel}.
Because the LAO are deposited to fit the substrate STO, the lattice constants of the deposited LAO layers in the direction horizontal to the interface are same as those of the substrate STO, while that in the direction perpendicular to the interface slightly changes depending on the carrier density at the interface.
Therefore, in this study, the structure is fixed as cubic, and the lattice constants for the bulk and the heterostructure perpendicular to the [$001$] stacking direction are fixed to $3.905$\AA , which corresponds to the experimental lattice constant of the bulk SrTiO$_3$. 
In the LDA calculations, $8 \times 8 \times  8$ $k$-point sampling is employed for the bulk, and $8 \times 8 \times  2$ $k$-point sampling is employed for the LAO/STO to represent electronic structures of the system.
The muffin-tin (MT) radii are: $R_{\text{Ti}} ^{\text{MT}} = 2.50$ bohr, $R_{\text{Sr}} ^{\text{MT}}=2.10 $bohr, $R_{\text{Al}} ^{\text{MT}} =1.90$ bohr, $R_{\text{La}} ^{\text{MT}} =1.60$ bohr, and $R_{\text{O}} ^{\text{MT}} =1.6$ bohr.
The angular momentum cut off is taken at $l=4$ for all the sites.
In the cRPA and the GW calculations, $3 \times 3 \times  3$ $k$-point sampling is employed for the bulk, and $3 \times 3 \times  1$ $k$-point sampling is employed for the LAO/STO.


\section{Results}
\subsection{band structure and density of state}

\begin{figure*}[htb]
\begin{center} 
\includegraphics[width=0.9\textwidth ]{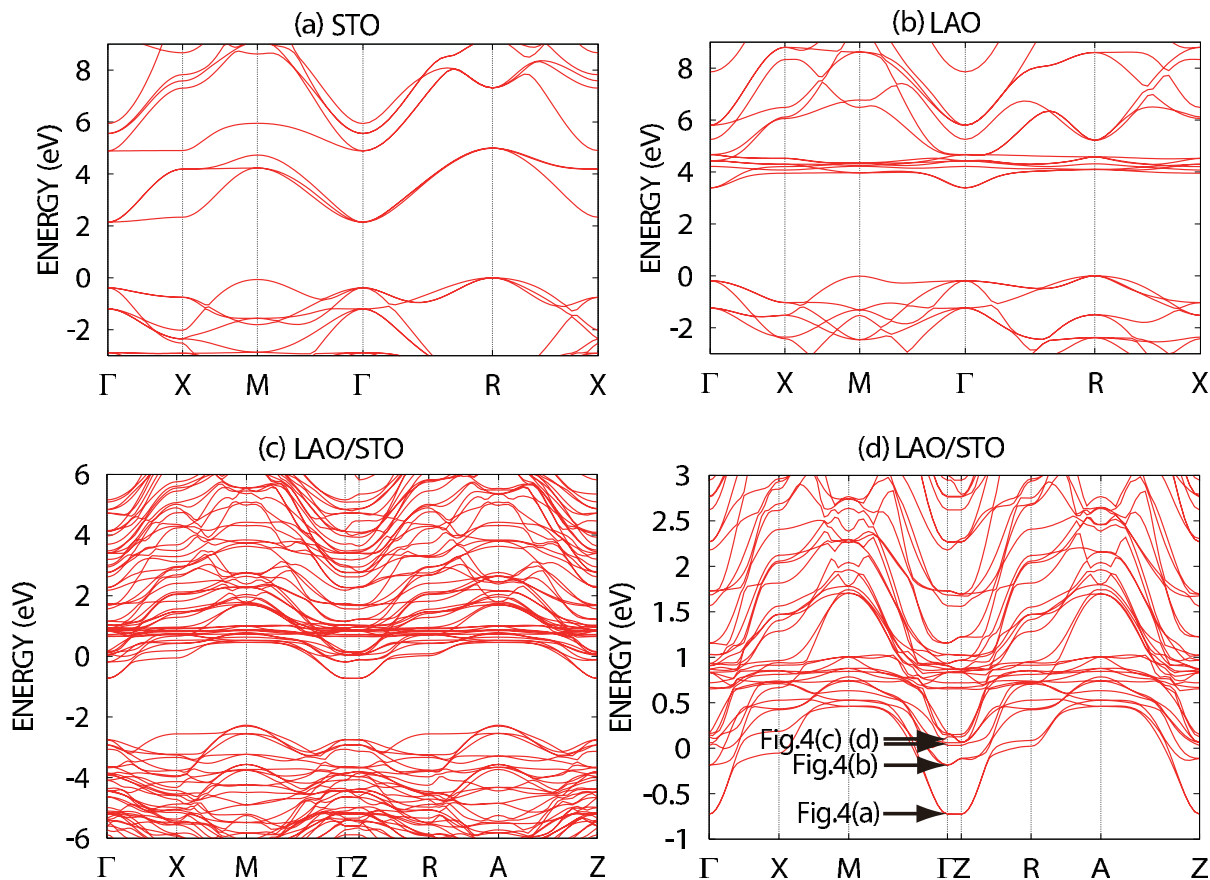} 
\end{center} 
\caption{ Electronic band structures obtained by the LDA. The zero energy corresponds to the Fermi level. Upper left panel:  Electronic band of STO. Upper right panel:  Electronic band of LAO. 
Lower panels: Electronic bands of LAO$1.5$STO$3.5$. 
The right one is the enlarged view around the Fermi level. 
Wave functions of the states indicated by arrows are displayed in Fig. \ref{BlochLAOSTO}.}
\label{bndsBULKandLAOSTO}
\end{figure*} 
\begin{figure}[tb]
\centering
\includegraphics[clip,width=0.45\textwidth ]{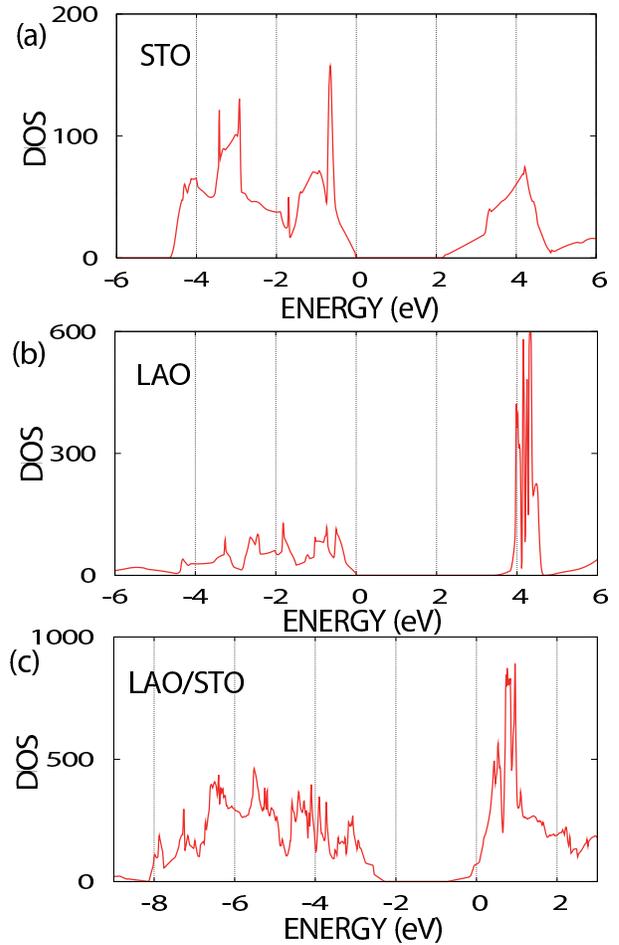} 
\caption{(color online) Density of states of bulk STO and LAO, and LAO/STO obtained by LDA. Energy is measured from the Fermi level.}
\label{dostot}
\end{figure}
\begin{figure}[tb]
\centering
\includegraphics[clip,width=0.3\textwidth ]{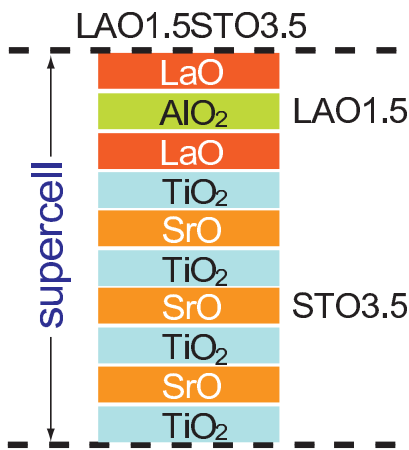} 
\caption{(color online) Schematic picture of the model of the transition-metal-oxide heterostructure LAO$1.5$STO$3.5$. The system is modeled by a periodically repeated supercell containing two interfaces. 
The both interfaces are LaO-TiO$_2$. 
Owing to neutrality of the electrons and atoms, these systems are formally charged by $-e$ electron per supercell.}
\label{supercell}
\end{figure}

First, we show the band structure and the density of states of the insulators, bulk SrTiO$_3$ and LaAlO$_3$.
The upper panels of Figs. \ref{bndsBULKandLAOSTO} and \ref{dostot} show the band structure and the density of states , respectively.
At room temperature, SrTiO$_3$ has the cubic perovskite structure, and becomes tetragonal below $105$ K. 
SrTiO$_3$ has a high dielectric constant at low temperatures because of the nature of quantum paraelectricity\cite{mullar}.   
In this calculation, the structure is fixed as cubic, and the lattice parameters are fixed at $3.905$ \AA \ of the bulk SrTiO$_3$.
The lower three conduction bands are derived from the $t_{2g}$ orbital of Ti sites, where the octahedral crystal field of O$^{2-}$ partially breaks the $5$-fold symmetry of the $3d$ orbitals into the lower orbitals of the $t_{2g}$ and the higher orbital $e_g$.
The calculated band gap is $2.1$ eV and the band width of the $t_{2g}$ band is $2.8$ eV.
In the experiment, the band gap is $3.3$ eV\cite{benthem}.
In the strongly correlated materials, such underestimation of the band gap causes to overestimate the screening effect from the high-energy bands to the low-energy bands in the cRPA.  
In the LAO/STO, the underestimation of the energy levels of La $4f$ bands is a major problem which are entangled with Ti $t_{2g}$ bands in the LDA as we will show.
LaAlO$_3$ has the rhombohedral perovskite structure at room temperature, and becomes cubic above $821$ K.
To compare with the LAO layer in the heterostructure, the structure and the lattice parameters are fixed at those of the cubic SrTiO$_3$ in this calculation.
The lower narrow conduction bands of LAO at $\sim$4 eV are derived from the $4f$ orbital of La sites.
The La $5d$ bands are hybridized with the La $4f$ bands  at the $\Gamma $-point.  
In the experiment, the band gap between the La $4f$ and the O $2p$ is $5.6$ eV \cite{lim}. 
In this calculation, however, the band gap is $3.4$ eV because the LDA underestimates 
the value of the band gap.
This band-gap problem in the LDA is improved dramatically with the GWA.
We will show a result in the GWA later.

\begin{figure}[tb]
\centering
\includegraphics[clip,width=0.45\textwidth ]{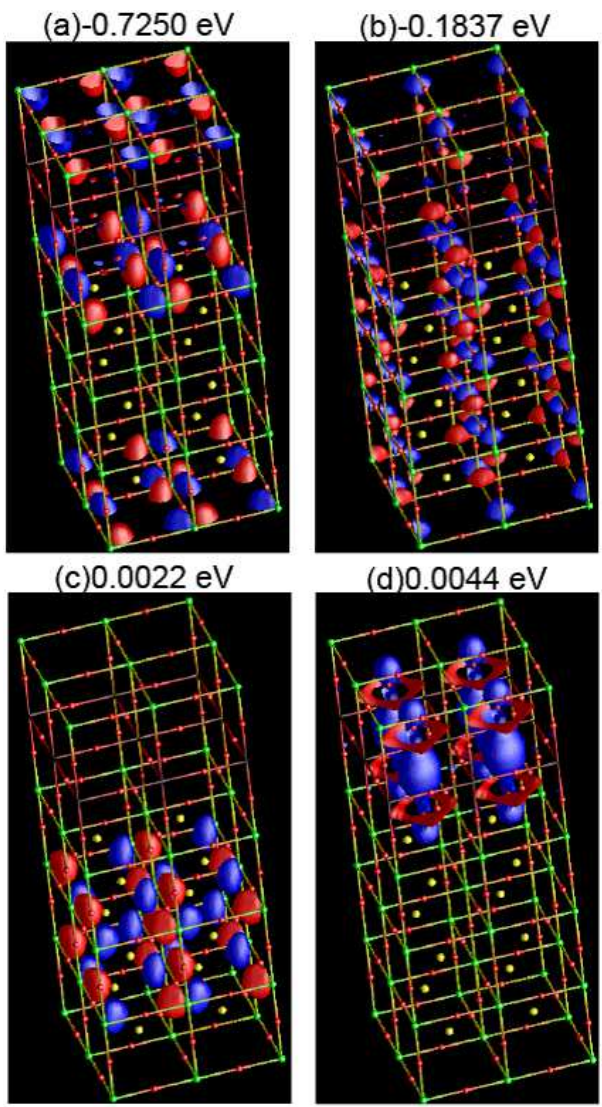} 
\caption{(color online) Isosurface contours of the Bloch functions of LAO$1.5$STO$3.5$ conduction bands at $\Gamma$-point [$000$]. The structure corresponds to Fig. \ref{supercell}, for example, the panel A represents the Ti $d_{xy}$ at the interface. Value of the isosurface is set to $\pm 0.05$  bohr$^{-3/2}$. Each energy level is measured from the Fermi level and is depicted as arrows in Fig. \ref{bndsBULKandLAOSTO}. }
\label{BlochLAOSTO}
\end{figure}

Next, we show the LDA results of the transition-metal-oxide heterostructure LAO$1.5$STO$3.5$.
We refer to the TiO$_2$ layer at the interface as the $1$st-layer and the TiO$_2$ layer in the bulk region of the SrTiO$_3$ as the $2$nd-layer.   
In this paper, we refer to the heterostructure -(LaAlO$_3$)$_1$-LaO/TiO$_2$-(SrTiO$_3$)$_3$- as LAO$1.5$STO$3.5$ (see Fig. \ref{supercell}).   
This heterostructure has two crystallographically equivalent $n$-type interface, and has about $1/2$ carrier electron at each $n$-type interface, because LAO$1.5 ^{+1}$STO$3.5 ^{0}$ has a positive charge in the ionic limit.
In terms of the polar discontinuity, these models of the heterostructures are the cases where the instabilities of the potential divergence are completely suppressed\cite{nakagawa,hirayama}.
The lower panels of Fig. \ref{bndsBULKandLAOSTO} show the band structure of the $n$-type LAO$1.5$STO$3.5$ heterostructure.
The energy bands are rather degenerate because there are two crystallographically equivalent $n$-type interfaces.
The LAO$1.5$STO$3.5$ has the Fermi surface around the $\Gamma $ point.
The density of states of the LAO/STO comes from nearly superimposed states of the bulk LAO and STO (see Fig. \ref{dostot}).

\begin{figure}[tb]
\centering
\includegraphics[clip,width=0.45\textwidth ]{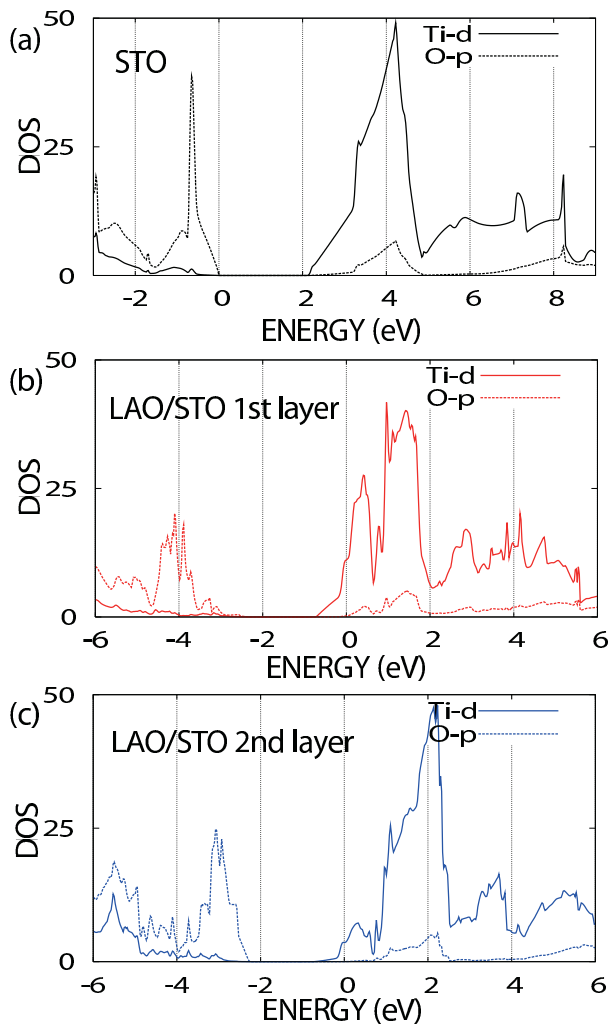} 
\caption{(color online) Partial densities of states of Ti $d$ and O $p$ at TiO$_2$ layers of bulk STO and LAO/STO obtained by the LDA. Energy is measured from the Fermi level.}
\label{dosTiO}
\end{figure}

Figures \ref{BlochLAOSTO} shows the isosurface contours of selected Bloch functions of the conduction bands at the $\Gamma$-point.
The panel (a) of Fig. \ref{BlochLAOSTO} is the isosurface contour of the lowest conduction band (see Fig. \ref{bndsBULKandLAOSTO} (d)).
This band is originated mainly from the Ti $d_{xy}$ in the $1$st-layer.
The panel (b) is the isosurface contour of the Ti $d_{yz}$ band, which spreads in the direction horizontal to the interface. 
The panel (c) and (d) are the isosurface contours of the Ti $d_{xy}$ band in the $2$nd-layer and the La $4f$ band, respectively.
The Ti $t_{2g}$ and La $4f$ orbitals are spatially close and, in the LDA level, energetically close, so that the Ti $t_{2g}$ and La $4f$ orbitals are hybridized at the interface (see Fig. \ref{BlochLAOSTO}). 
The Bloch functions transfer to the bulk region with increasing those energy levels.
There is a positive crystal field from the polar perovskite LaAlO$_3$ to the non-polar substrate SrTiO$_3$ as compared with the case with only Sr$^{2+}$ for the bulk SrTiO$_3$.
This crystal field vanishes in the bulk region of SrTiO$_3$, because the negative field from the doped electrons compensates this positive field.
From these reasons, the energy levels of the orbitals in the $1$st-layer are lower than that in the $2$nd-layer, and the gap between the valence and conduction bands in LAO$1.5$STO$3.5$ is smaller than that of the bulk STO. 
Such effect of compensation and confinement becomes strong as the polar perovskite LaAlO$_3$ becomes thicker.
This tendency is also seen in Fig. \ref{dosTiO} which shows the partial density of states at the Ti sites of $d$ orbitals and the O sites of $p$ orbitals in the TiO$_2$ layer of the bulk STO and LAO/STO obtained by the LDA.
Hereafter, for instance, a Wannier orbital of the $xy$ orbital in the $1$st-layer is denoted by $1xy$. 
Pronounced peak shifts of DOS to the lower energy are found for the sites in the $1$st-layer.

\begin{figure}[tb]
\centering
\includegraphics[clip,width=0.45\textwidth ]{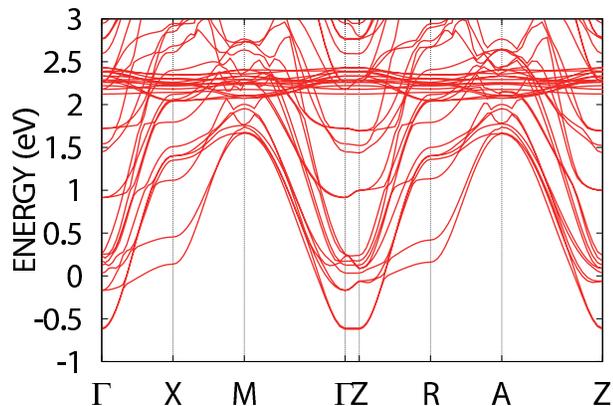} 
\caption{(color online) Electric band structures with $1$-shot GW self-energy for $4f$ bands. Energy is measured from the Fermi level.}
\label{bndsGWLAOSTO}
\end{figure}

Considering the experimental band gap of the bulk LaAlO$_3$, we note that the location of the energy of the La $4f$ bands in the LaAlO$_3$/SrTiO$_3$ is too low in the LDA.
The La $4f$ bands screen and hybridize with the Ti $t_{2g}$ bands weaker in the real material.
To calculate the parameter of the low-energy model with a high accuracy, a better way is to take into account the correlation effect of  La separately by the GWA. 
To calculate the self energies of the La $4f$ bands, we first construct the maximally localized Wannier function of La $4f$ from a linear combination of the target low-energy KS-bands.
We choose the energy window from $-1$ eV to $3$ eV for the Wannier functions.
We find, however, that the screened Coulomb interaction of the low-energy bands derived from the Ti $t_{2g}$ is not sensitive to the choice of the energy window if the window exceeds a certain width but is not too wide.
Next, we calculate the self-energy corrections of the La $4f$ bands by the $1$-shot GW scheme. 
Figure \ref{bndsGWLAOSTO} shows the band structure after the $1$-shot GW corrections for the $4f$ bands.
The energy level of $4f$ bands is raised by about $1.5$ eV with GW. 
This value is consistent with the experimental result of the LAO\cite{lim}.
In the following sections, we show this self-energy effect on the effective model parameters.


\subsection{Wannier function and transfer integral}

\begin{figure}[tb]
\centering
\includegraphics[clip,width=0.45\textwidth ]{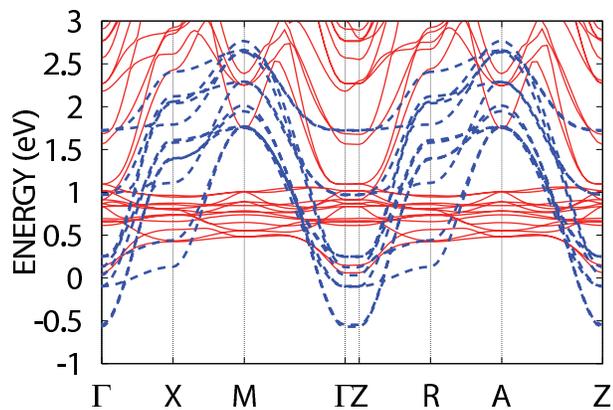} 
\caption{(color online) Disentangled $d$-bands having strong $t_{2g}$ character ((blue) dashed line) and diagonalized $r$-bands (solid line).}
\label{bndscRPALAOSTO}
\end{figure}
\begin{figure}[tb]
\centering
\includegraphics[clip,width=0.45\textwidth ]{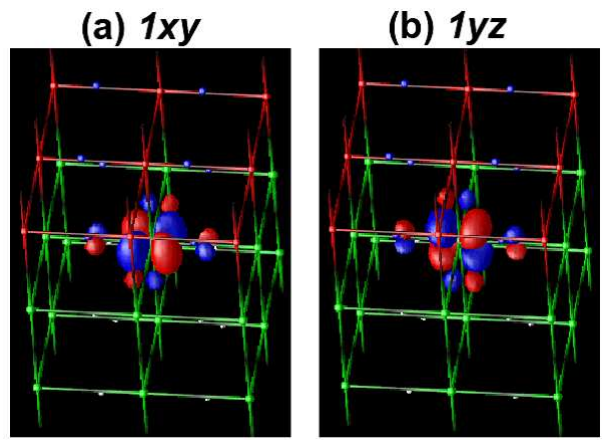} 
\caption{(color online) Isosurface of the maximally localized Wannier functions $\pm 0.05$ a.u. for the $d_{xy}$ and $d_{yz}$ orbitals in the $1$st-layer. }
\label{WannierLAOSTO}
\end{figure}

We calculate the transfer integral $t$ of the Ti $t_{2g}$ orbital to determine the $1$-body part of low-energy effective Hamiltonian for the heterostructure.
First, we construct $6\times 2$ maximally localized Wannier functions having strong Ti $t_{2g}$ characters, from the linear combination of the target low-energy KS-bands.
We choose $-1$-$3$ eV as the energy window for the Wannier functions.
Figure \ref{bndscRPALAOSTO} shows the disentangled $t_{2g}$ bands and the rest bands, and Fig. \ref{WannierLAOSTO} shows the isosurfaces of Wannier functions of the $d_{xy}$ and $d_{yz}$ orbitals.

We show in Table \ref{t_STO} the transfer integrals $t$ of the bulk STO calculated in the LDA.  
In the tables and this subsection, $m$ and $n$ specify symmetries of the Ti $t_{2g}$ orbitals. 
The values of on-site energies are listed in the column for $(R_{x},R_{y},R_{z}) = (0, 0, 0)$.
The values of the on-site energies are the same for all the $t_{2g}$ orbitals because of the cubic crystal symmetry. 
The major values of the nearest hopping  between the same symmetries $(R_x ,R_y ,R_z ) = (1, 0, 0)$ are $299$ meV.
These values are consistent with the values in the literature in nearest-neighbor tight-binding models of the transition-metal-oxide interfaces ($t\sim 0.3$ eV) \cite{okamoto}.
The nearest hopping for the perpendicular directions of the orbitals symmetry is $-39$ meV, which is $13$ percent of these of the main directions.  
The next nearest hopping $(R_{x},R_{y},R_{z}) = (1, 1, 0)$ is $35$ percent of the nearest neighbor hopping.

Next, we show in Table \ref{t_LDA} the transfer integrals $t$ of the LAO$1.5$STO$3.5$ calculated in the LDA and the GW, respectively.  
The values of on-site energy in the $1$st-layer $t_{1m,1m} (0, 0, 0)$ are about $650$ meV lower than those in the $2$nd-layer $t_{2m,2m} (0, 0, 0)$, mainly because the dipole moment of LaO$^{1+}$-AlO$_{2} ^{1-}$ stabilizes the energy of the $t_{2g}$ orbitals in the $1$st-layer.
The hybridization between the La $4f$ and Ti $t_{2g}$ orbitals also slightly stabilizes the energy of the $t_{2g}$ orbitals.
These layer dependent potential localizes carriers at the interface.
Actually, in the experiment, the transition into the $2$D superconducting state, namely the Berezinskii-Kosterlitz-Thouless transition is seen at the interface of the LAO/STO\cite{reyren}.  
The value of the on-site energy of $1xy$, $t_{1xy,1xy} (0, 0, 0)$, is lower than that of the other $1t_{2g}$ due to the crystal field and the hybridization with the LAO layer.
In the GWA, this difference of the on-site energy at the $1$st-layer is $16$ meV larger than that in the LDA, because the hybridizations of the La $4f$ orbitals with the Ti $t_{2g}$ orbitals in the $1$st-layer, especially with $1yz$ and $1zx$, become weaker.
In the $2$nd-layer, the dipole moment of LaO$^{1+}$-AlO$_{2} ^{1-}$ have little effect, and the values of on-site energy partially recover the bulk STO nature.
The hopping parameters are similar to those of the bulk STO.
The major values of the nearest hopping between the same symmetries $t_{1m,1m} (1, 0, 0)$ are about $0.3$ eV.
The nearest hopping of the $1xy$ is smaller than that of the bulk STO.
The nearest hopping of the $1yz$ and $1zx$ are nearly the same as those of the bulk STO.   
The main difference of the hopping parameters from the bulk STO is seen in the nearest hopping for the perpendicular directions of the orbitals symmetry $t_{1yz,1yz} (1, 0, 0)$.
Such hoppings are $-59$ meV, which are about twice as large as those of the bulk STO.
In the LAO/STO, therefore, the carriers tend to spread in the direction horizontal to the interface compared to the bulk STO.
As with the on-site energy, the hopping parameters in the $2$nd-layer $t_{2m,2m} (1, 0, 0)$ are nearly the same as those of the bulk STO.
In the GWA, the hybridization between the La $4f$ and $t_{2g}$ orbitals in the $1$st-layer, especially $1yz$ and $1zx$, becomes weaker.
As a result, the splitting of on-site energies of $t_{2g}$ orbitals becomes larger, and the hopping parameter between $1zx$-$2zx$ becomes smaller.

\begin{table*}[htb] 
\caption{Transfer integrals for the $t_{2g}$ orbitals of the Ti sites in the bulk STO, $t_{mn}(R_x, R_y, R_z)$, 
where $m$ and $n$ denote symmetries of $t_{2g}$ orbitals. Units are given in meV.  
} 
\
\begin{tabular}{c|rrr} 
\hline \hline \\ [-4pt]
STO \\ [+2pt] 
\hline \\ [-4pt]
\backslashbox{$(m, n)$}{$\bm{R}$} 
& \big[0,0,0\big] 
& \big[1,0,0\big] 
& \big[1,1,0\big] \\ [+4pt]
\hline \\ [-8pt]
$(xy,xy)$&   2430 & $-$299 & $-$109 \\
$(xy,yz)$&      0 &      0 &      0 \\
$(xy,zx)$&      0 &      0 &      0 \\ 
$(yz,yz)$&   2430 &  $-$39 &      7 \\
$(yz,zx)$&      0 &      0 &      9 \\ 
$(zx,zx)$&   2430 & $-$299 &      7 \\ 
\hline 
\hline 
\end{tabular}
\label{t_STO} 
\end{table*} 

\begin{table*}[htb] 
\caption{Transfer integrals for $t_{2g}$ orbitals of Ti sites in LAO$1.5$STO$3.5$ calculated in LDA and GWA level, $t_{mn}(R_x, R_y, R_z)$, 
where $m$ and $n$ denote layers and symmetries of $t_{2g}$ orbitals. Units are given in meV.  
The parameters calculated from the LDA band are listed as ``LDA'', and the parameters calculated from the LDA band with the self-energy correction of the La $4f$ are listed as ``GW''.
} 
\
\begin{tabular}{c|rrr|rrr} 
\hline \hline \\ [-4pt]
LAO/STO
&
&LDA 
&
& 
&GW
& \\ [+2pt] 
\hline \\ [-4pt]
\backslashbox{$(m, n)$}{$\bm{R}$} 
& \big[0,0,0\big] 
& \big[1,0,0\big] 
& \big[1,1,0\big] 
& \big[0,0,0\big] 
& \big[1,0,0\big] 
& \big[1,1,0\big] \\ [+4pt]
\hline \\ [-8pt]
$(1xy,1xy)$&   2606 & $-$289 & $-$101 &   2607 & $-$288 & $-$104 \\
$(1xy,1yz)$&      0 &   $-$7 &   $-$1 &      0 &   $-$4 &      2 \\
$(1xy,1zx)$&      0 &      0 &   $-$1 &      0 &      0 &      2 \\ 
$(1xy,2xy)$&  $-$42 &      6 &   $-$9 &  $-$42 &      6 &   $-$9 \\ 
$(1xy,2yz)$&      0 &     10 &      6 &      0 &      9 &      6 \\
$(1xy,2zx)$&      0 &      0 &      6 &      0 &      0 &      6 \\ 
$(1yz,1yz)$&   2617 &  $-$57 &   $-$2 &   2634 &  $-$59 &   $-$5 \\
$(1yz,1zx)$&      0 &      0 &   $-$6 &      0 &      0 &   $-$9 \\ 
$(1yz,2xy)$&      0 &      8 &      8 &      0 &      8 &      8 \\
$(1yz,2yz)$& $-$296 &      3 &   $-$6 & $-$299 &      4 &   $-$5 \\
$(1yz,2zx)$&      0 &      0 &      8 &      0 &      0 &      8 \\ 
$(1zx,1zx)$&   2617 & $-$302 &   $-$2 &   2634 & $-$298 &   $-$5 \\ 
$(1zx,2xy)$&      0 &      0 &      8 &      0 &      0 &      8 \\ 
$(1zx,2yz)$&      0 &      0 &      8 &      0 &      0 &      8 \\ 
$(1zx,2zx)$& $-$296 &  $-$98 &   $-$6 & $-$299 &  $-$98 &   $-$5 \\ 
$(2xy,2xy)$&   3268 & $-$298 & $-$109 &   3264 & $-$298 & $-$109 \\
$(2xy,2yz)$&      0 &   $-$3 &      0 &      0 &   $-$3 &      0 \\
$(2xy,2zx)$&      0 &      0 &      0 &      0 &      0 &      0 \\ 
$(2yz,2yz)$&   3262 &  $-$39 &      6 &   3262 &  $-$39 &      6 \\
$(2yz,2zx)$&      0 &      0 &      8 &      0 &      0 &      7 \\ 
$(2zx,2zx)$&   3262 & $-$297 &      6 &   3262 & $-$297 &      6 \\ 
\hline 
\hline 
\end{tabular}
\label{t_LDA} 
\end{table*}

\subsection{screened Coulomb interaction}

Next, we calculate the on-site screened Coulomb interaction $U_{mn}(0, 0, 0)$ and the on-site screened exchange interaction $J_{mn}(0, 0, 0)$ of the Ti $t_{2g}$ orbital to determine the $2$-body part of the low-energy effective Hamiltonian for the heterostructure.   
The screened Coulomb interaction is computed with the matrix elements in the maximally localized Wannier basis in the framework of the cRPA (see eq. (\ref{Wrrep})).

In the top of Table \ref{W_STOLDA}, we show the effective on-site Coulomb interaction $U$ and $J$ for the bulk STO. 
As with the transfer integrals, $m$ and $n$ denote symmetries of the $t_{2g}$ orbitals of Ti in the tables and this subsection. 
The effective on-site Coulomb interaction $U$ between the same orbitals is $3.76$ eV, while the bare on-site Coulomb interaction is $14.27$ eV.
The on-site Coulomb interaction is reduced about to $1/4$ by the polarizations without the $d$-$d$ contributions. 
The effective on-site screened exchange interaction $J$ is $0.46$ eV.

We show the effective on-site Coulomb interaction $U$ and $J$ of LAO/STO calculated with cRPA in the middle and bottom of the Table \ref{W_STOLDA}.      
Here, $U$ between the same orbitals are $3.4$-$3.6$ eV from the LDA and $3.7$-$4.0$ eV from the GWA, while that of the bulk STO is $3.76$ eV. 
In both the LDA and GWA, $U_{1xy,1xy}$ is the largest among on-site screened Coulomb interactions in the $1$st-layer, and $U_{1yz,1yz}$ and $U_{1zx,1zx}$ are $4$-$5$ percent smaller than $U_{1xy,1xy}$.   
In the GWA, $U$ and $J$ at the $1$st-layer become larger than those of the bulk STO, while these are smaller in the LDA.
This is because the La bands are raised away from the Fermi level by the self-energy correction, the polarizations between La $4f$ and Ti $t_{2g}$ orbitals become weaker compared to the LDA bands, and, as a result, the screening effect becomes weaker.  
The on-site screened Coulomb interaction $U_{mn}$ and the on-site screened exchange interaction $J_{mn}$ satisfy the equation $U_{mn}=U_{mm}-2J_{mn}$ in the bulk STO.
On the other hand, at the interface of the LAO/STO, these parameters do not satisfy this equation because of the inversion symmetry breaking. 
In the $2$nd-layer, $U$ and $J$ recover the bulk STO nature.   
Although, similarly in the $1$st-layer, $U_{2xy,2xy}$ is the largest on-site screened Coulomb interaction in the $2$nd-layer, the difference between $U_{2xy,2xy}$ and the others of $t_{2g}$ is only about $1$ percent.
Because the electrons confined at the interface have low dimensionality, the correlation becomes effectively stronger than the bulk STO, even if the transfer integrals at the interface are similar to those of the bulk STO.

\begin{table*} 
\caption{
Effective Coulomb interaction ($U$)/exchange ($J$) interactions between the two electrons for all the combinations of Ti-$t_{2g}$ orbitals in the bulk STO and LAO$1.5$STO$3.5$, respectively (in eV). Especially, for the combinations between the orbitals in the same Ti sites, the matrix elements represent the effective on-site Coulomb interaction. The band structures of LAO$1.5$STO$3.5$ are calculated in the LDA and GWA.
}
\ 
\label{W_STOLDA} 
\begin{tabular}{ccccccccccccccccc} 
\hline \hline \\ [-8pt]  
STO &  &      &      & $U$  &      &      & &       &      &      &      &  $J$ &      &      \\ [+1pt]
\hline \\ [-8pt] 
      & $xy$ & $yz$ & $zx$ &      &      &      & &       & $xy$ & $yz$ & $zx$ &      &      &      \\ 
\hline \\ [-8pt] 
$xy$  & 3.76 & 2.81 & 2.81 &      &      &      & & $xy$  &      & 0.46 & 0.46 &      &      &      \\ 
$yz$  & 2.81 & 3.76 & 2.81 &      &      &      & & $yz$  & 0.46 &      & 0.46 &      &      &      \\
$zx$  & 2.81 & 2.81 & 3.76 &      &      &      & & $zx$  & 0.46 & 0.46 &      &      &      &      \\
\hline \hline \\ [-8pt]
LAO/STO(LDA) & & & & $U$ &   &      & &       &      &      &      & $J$  &      &      \\ [+1pt]
\hline \\ [-8pt] 
      & $1xy$ & $1yz$ & $1zx$ & $2xy$ & $2yz$ & $2zx$ & & & $1xy$ & $1yz$ & $1zx$ & $2xy$ & $2yz$ & $2zx$ \\ 
\hline 
$1xy$ & 3.65 & 2.63 & 2.63 & 0.49 & 0.55 & 0.55 & & $1xy$ &      & 0.46 & 0.46 & 0.00 & 0.00 & 0.00 \\ 
$1yz$ & 2.63 & 3.48 & 2.58 & 0.55 & 0.65 & 0.63 & & $1yz$ & 0.46 &      & 0.44 & 0.00 & 0.01 & 0.00 \\
$1zx$ & 2.63 & 2.58 & 3.48 & 0.55 & 0.63 & 0.65 & & $1zx$ & 0.46 & 0.44 &      & 0.00 & 0.00 & 0.01 \\
$2xy$ & 0.49 & 0.55 & 0.55 & 3.62 & 2.67 & 2.67 & & $2xy$ & 0.00 & 0.00 & 0.00 &      & 0.46 & 0.46 \\ 
$2yz$ & 0.55 & 0.65 & 0.63 & 2.67 & 3.59 & 2.65 & & $2yz$ & 0.00 & 0.01 & 0.00 & 0.46 &      & 0.46 \\ 
$2zx$ & 0.55 & 0.63 & 0.65 & 2.67 & 2.65 & 3.59 & & $2zx$ & 0.00 & 0.00 & 0.01 & 0.46 & 0.46 &      \\
\hline \hline \\ [-8pt]  
LAO/STO(GW)  & & & & $U$ & & & &   &      &    &  &  $J$ & &   \\ [+1pt]
\hline \\ [-8pt] 
      & $1xy$ & $1yz$ & $1zx$ & $2xy$ & $2yz$ & $2zx$ &  &  & $1xy$ & $1yz$ & $1zx$ & $2xy$ & $2yz$ & $2zx$ \\  
\hline \\ [-8pt] 
$1xy$ & 4.00 & 2.98 & 2.98 & 0.72 & 0.79 & 0.79 & & $1xy$ &      & 0.46 & 0.46 & 0.00 & 0.00 & 0.00 \\ 
$1yz$ & 2.98 & 3.83 & 2.93 & 0.78 & 0.89 & 0.87 & & $1yz$ & 0.46 &      & 0.44 & 0.00 & 0.01 & 0.00 \\
$1zx$ & 2.98 & 2.93 & 3.83 & 0.78 & 0.87 & 0.89 & & $1zx$ & 0.46 & 0.44 &      & 0.00 & 0.00 & 0.01 \\
$2xy$ & 0.72 & 0.78 & 0.78 & 3.80 & 2.84 & 2.84 & & $2xy$ & 0.00 & 0.00 & 0.00 &      & 0.46 & 0.46 \\ 
$2yz$ & 0.79 & 0.89 & 0.87 & 2.84 & 3.76 & 2.83 & & $2yz$ & 0.00 & 0.01 & 0.00 & 0.46 &      & 0.46 \\ 
$2zx$ & 0.79 & 0.87 & 0.89 & 2.84 & 2.83 & 3.76 & & $2zx$ & 0.00 & 0.00 & 0.01 & 0.46 & 0.46 &      \\
\hline 
\hline
\end{tabular} 
\end{table*}


\section{Summary} 
In this paper, we have determined the parameters of the low-energy effective model of LaAlO$_3$/SrTiO$_3$ by the MACE.
As with many interfaces, LaAlO$_3$/SrTiO$_3$ has a complex band structure where the bands of both the interface and bulk regions are highly entangled with each other.    
By the disentangling scheme using the maximally localized Wannier function\cite{miyake3}, we disentangle the low-energy part having the strong characters of Ti $t_{2g}$ orbital from the global KS-bands, and thus enable to calculate the parameters in the effective Hamiltonian of LaAlO$_3$/SrTiO$_3$ by the cRPA.
The parameters in this study offer not only the superlattice model but also the semi-infinite interface model (vacuum)-(LaAlO$_3$)$_{N}$-(SrTiO$_3$)$_{\infty}$, because the screening effect from the inter-supercell rapidly decreases with distance and thus the superlattice model corresponds to the semi-infinite interface model in the limit of thick LAO and STO.
The parameters have anisotropies and a layer dependence.
The on-site energies in the $1$st-layer are $650$ meV higher than those in the $2$nd-layer, which causes localization of the carriers at the interface.  
In the $1$st-layer, while the transfer integral of the $t_{2g}$ orbital is similar to that of the bulk SrTiO$_3$, the screened Coulomb interaction $U$ of the $t_{2g}$ orbital becomes $10$ percent larger than that of the bulk SrTiO$_3$.
In the bulk region of LaAlO$_3$/SrTiO$_3$, the parameters of the $t_{2g}$ orbitals recover the values and the symmetry of the bulk SrTiO$_3$.
The obtained $3$-dimensional parameters constitute the low-energy effective models of LaAlO$_3$/SrTiO$_3$, either in semi-infinite structures in one of the directions or in supercell structures.
The resultant low-energy effective model offers a firm and quantitative basis when one wishes to solve the effective model by using accurate low-energy solvers in the future.

Recently, an \textit{ab-initio} dimensional downfolding scheme, which downfolds a $3$-dimensional model to a lower-dimensional model in real space, has been formulated\cite{nakamura3}. 
By applying the dimensional downfolding scheme to the $3$-dimensional model of LAO/STO, we can also obtain a $2$-dimensional model of a single LAO/STO supercell. 
In the dimensional downfolding, one is able to expect that the weakness of inter-layer (inter-chain) couplings justify the RPA type perturbative treatment.\cite{imada}
In the present case, however, the interface layer is not particularly weakly coupled with the other layers.
Nevertheless, the layers far away from the interface do not join in the low-energy excitations if the bulk is insulating as in the present case.
Therefore, the gapped excitations in the bulk part can be safely downfolded into the metallic low-energy excitation near the interface in the same spirit of the cRPA.
This enables to derive effective low-energy models solely for the interface part.
Of course, we may need to keep the bands of not one but several layers near the interface in the low-energy models.
This dimensional downfolding is a challenging future issue.

    
\acknowledgment
MH would like to thank Kazuma Nakamura, Takahiro Misawa, Youhei Yamaji, Hiroshi Shinaoka, and Ryota Watanabe for useful advices and fruitful discussions.
This work has been supported by Grant-in-Aid for Scientific Research from MEXT Japan under the grant numbers 22104010 and 22340090. This work has also been financially supported by MEXT HPCI Strategic Programs for Innovative Research (SPIRE) and Computational Materials Science Initiative (CMSI).

    
\appendix
\section{GWA for disentangled band}

The quasiparticle energies and wave functions are obtained by solving
\begin{equation}
\begin{split}
(T+V_{\text{ext}}+V_{\text{H}})\psi _{nk}(\bm{r})+\int d\bm{r'}\Sigma &(\bm{r},\bm{r'};E_{n\bm{k}})\psi _{n\bm{k}}(\bm{r'})\\
                                                 &=E_{n\bm{k}}\psi _{n\bm{k}}(\bm{r}), 
\label{sigmaEnondia}
\end{split}
\end{equation}
where $T$ is the kinetic energy operator, and $V_{\text{ext}}$ is the external potential, $V_{\text{H}}$ is the Hartree potential.
If we consider only the diagonal parts of the self-energy $\Sigma$, eq. (\ref{sigmaEnondia}) is reduced to
\begin{equation}
E _{n\bm{k}} =\epsilon _{n\bm{k}} ^{\text{LDA}} -\langle   n\bm{k}|V _{\text{ex}} ^{\text{LDA}}|n\bm{k} \rangle +\langle  n\bm{k}|\Sigma (E_ {n\bm{k}} )|n\bm{k}\rangle .
\label{sigmaEdia}
\end{equation}
The self-energy operator must be estimated at the quasiparticle energy $E _{n\bm{k}}$.
This is done by expanding the matrix elements of the self-energy operator to the first order in the energy around $\epsilon _{n\bm{k}}$.
Then the quasiparticle energy is obtained explicitly;
\begin{equation}
E _{n\bm{k}} =\epsilon _{n\bm{k}} +Z_{n\bm{k}}(\epsilon _{n\bm{k}})(\epsilon _{n\bm{k}} ^{\text{LDA}} -\epsilon _{n\bm{k}}+ \Delta \Sigma _{n\bm{k}}(\epsilon _{n\bm{k}})),
\label{qpEep}
\end{equation}
where the self-energy is given by
\begin{equation}
\Delta \Sigma _{n\bm{k}}=\Sigma _{n\bm{k}} -V_{\text{ex}} ^{\text{LDA}}
\label{DeSi}
\end{equation} 
and $Z_{n\bm{k}}$ is the renormalization factor
\begin{equation}
Z_{n\bm{k}}=(1-\frac{\partial \Delta \Sigma _{n\bm{k}}(\epsilon _{n\bm{k}})}{\partial \omega})^{-1}.
\label{Zfac}
\end{equation} 
If $\epsilon _{n\bm{k}}$ coincides with $\epsilon _{n\bm{k}} ^{\text{LDA}}$, then eq. (\ref{qpEep}) is simplified;  
\begin{equation}
E _{n\bm{k}} =\epsilon _{n\bm{k}} ^{\text{LDA}} +Z_{n\bm{k}}(\epsilon _{n\bm{k}} ^{\text{LDA}})\Delta \Sigma _{n\bm{k}}(\epsilon _{n\bm{k}} ^{\text{LDA}}).
\label{qpELDA}
\end{equation}

\begin{figure}[tb]
\centering
\includegraphics[clip,width=0.45\textwidth ]{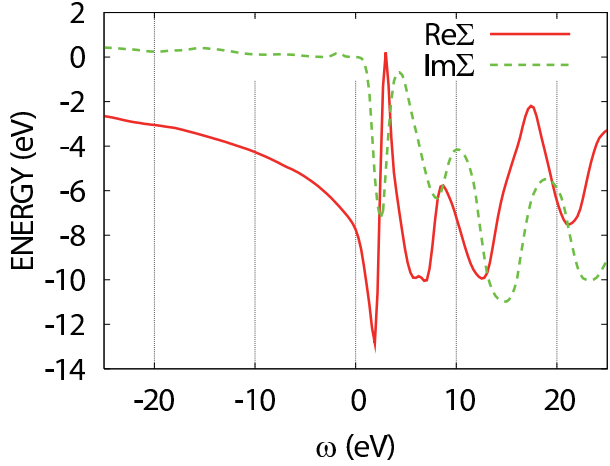} 
\caption{Real and imaginary parts of the matrix elements of the self-energy $\Sigma $ for a $4f$ Wannier band in LAO$1.5$STO$3.5$ at $\Gamma $-points. Solid and dashed lines represent the real and imaginary parts of the $\Sigma $, respectively.  }
\label{Sigma}
\end{figure}

We show the real and imaginary parts of the self-energy for a $4f$ Wannier band calculated by the $1$-shot GW approximation in Fig. \ref{Sigma}. 
The real and imaginary parts of the self-energy strongly oscillate above the Fermi level $E_{\text{F}}=0$ eV, especially at $2$-$3$ eV near the La $4f$ bands in the LDA level.
This lack of smoothness of the self-energy $\Sigma$ is obtained in the ``$1$-shot'' correction.
The self-energy $\Sigma$ should hopefully be calculated in a self-consistent procedure of the Hedin's set of coupled equations \cite{hedin}.
In this calculation, we have calculated the self-energy correction by one shot without iteration to save the computational cost.
Considering the unstable behavior of the self-energy near the La $4f$ bands, we expand eq. (\ref{sigmaEdia}) around $E_{\text{F}}=0$ eV and then quasiparticle energies are approximated in the first order as follows;  
\begin{equation}
E _{n\bm{k}} =Z_{n\bm{k}}(0)(\epsilon _{n\bm{k}} ^{\text{LDA}} +\Delta \Sigma _{n\bm{k}}(0)).
\label{qpE0}
\end{equation}
The approximation eq. (\ref{qpELDA}) is equivalent to eq. (\ref{qpE0}), which is justified if the linearity of self-energy $\Sigma $ in the low-energy region is eventually recovered in the self-consistent accurate estimate.
The smoothness of $\Sigma $ should be eventually obtained after the self-consistent calculation of the GWA.   
In this calculation, we average $Z_{n\bm{k}}(0)$ and $\Delta \Sigma _{n\bm{k}}(0)$ in terms of the band index and the $k$-points;  
\begin{equation}
E _{n\bm{k}} =\langle Z_{n\bm{k}}(0) \rangle _{n\bm{k}}(\epsilon _{n\bm{k}} ^{\text{LDA}} +\langle \Delta \Sigma _{n\bm{k}}(0) \rangle _{n\bm{k}}).
\label{qpE0av}
\end{equation}

Since we are interested in the La 4f level measured from the Fermi level, 
the self-energy correction to the La 4f level is corrected by subtracting the correction to the Fermi level. 
The latter is approximately evaluated by the GW self-energy for  
the bottom of the $t_{2g}$ band using eq.(\ref{qpE0}).
Figure \ref{bndsGWLAOSTO} shows the band structure thus obtained after the $1$-shot GW corrections for the $4f$ bands.


\begin{thebibliography}{99} 
\bibitem{ohtomo} A. Ohtomo and H. Y. Hwang: Nature (London) \textbf{427} (2004) 423.
\bibitem{thiel} S. Thiel, G. Hammer, A. Schmehl, C. W. Schneider, and J. Mannhart: Science \textbf{313} (2006) 1942.
\bibitem{nakagawa} N. Nakagawa, H. Y. Hwang, and D. A. Muller: Nat. Mater. \textbf{5} (2006) 204.
\bibitem{hirayama} M. Hirayama and M. Imada:  J. Phys. Soc. Jpn. \textbf{79} (2010) 034704.
\bibitem{reyren} N. Reyren, S. Thiel, A. D. Caviglia, L. Fitting Kourkoutis, G. Hammerl, C. Richter, C. W. Schneider, T. Kopp, A.-S. R$\ddot{\text{u}}$etschi, D. Jaccard, M. Gabay, D. A. Muller, J.-M. Triscone, and J. Mannhart: Science \textbf{317} (2007) 1196.
\bibitem{brinkman} A. Brinkman, M. Huijben, M. V. Zalk, J. Huijben, U. Zeitler, J. C. Maan, W. G. Van Der Wiel, G. Rijders, D. H. A. Blank, and H. Hilgenkamp: Nat. Mater. \textbf{6}  (2007) 493.
\bibitem{park}M. S. Park, S. H. Rhim, and A. J. Freeman: Phys. Rev. B \textbf{74} (2006) 205416. 
\bibitem{ishibashi} S. Ishibashi and K. Terakura: J. Phys. Soc. Jpn. \textbf{77}  (2008) 104706.
\bibitem{pentcheva}R. Pentcheva and W. E. Pickett: Phys. Rev. B \textbf{74} (2006) 035112.
\bibitem{imada} M. Imada, and T. Miyake: J. Phys. Soc. Jpn. \textbf{79} (2010) 112001.
\bibitem{tahara}D. Tahara and M. Imada: J. Phys. Soc. Jpn. \textbf{77} (2008) 114701.
\bibitem{metzner}W. Metzner and D. Vollhardt: Phys. Rev. Lett. 62 (1989) 324.
\bibitem{georges}A. Georges, G. Kotliar, W. Krauth, and M. J. Rosenberg: Rev. Mod. Phys. \textbf{68} (1996) 13.
\bibitem{imada2}M. Imada, and T. Kashima: J. Phys. Soc. Jpn. \textbf{69} (2000) 2723.
\bibitem{mizusaki}T. Mizusaki and M. Imada: Phys. Rev. B \textbf{74} (2006) 014421.
\bibitem{nakamura1}K. Nakamura, Y. Yoshimoto, R. Arita, S. Tsuneyuki, and M. Imada: Phys. Rev. B \textbf{77} (2008) 195126.
 
\bibitem{solovyev1}I. V. Solovyev, and M. Imada: Phys. Rev. B \textbf{71} (2005) 045103. 
\bibitem{solovyev2}I. V. Solovyev: Phys. Rev. B \textbf{73} (2006) 155117.
\bibitem{miyake} T.Miyake, and F. Aryasetiawan: Phys. Rev. B \textbf{77} (2008) 085122.
\bibitem{pavarini}E. Pavarini, S. Biermann, A. Poteryaev, A. I. Lichtenstein, A. Georges, and O. K. Andersen: Phys. Rev. Lett. \textbf{92} (2004) 176403.
\bibitem{imai1}Y. Imai, I. V. Solovyev, and M. Imada: Phys. Rev. Lett. \textbf{95} (2005) 176405.
\bibitem{imai2}Y. Imai and M. Imada: J. Phys. Soc. Jpn. \textbf{75} (2006) 094713.
\bibitem{otsuka}Y. Otsuka and M. Imada: J. Phys. Soc. Jpn. \textbf{75} (2006) 124707. 
\bibitem{nakamura4}K. Nakamura, Y. Yoshimoto, T. Kosugi, R. Arita, and M. Imada: J. Phys. Soc. Jpn. \textbf{78} (2009) 083710.
\bibitem{nakamura2} K. Nakamura, R. Arita, and M. Imada: J. Phys. Soc. Jpn. \textbf{77} (2008) 093711.
\bibitem{miyake2} T. Miyake, K. Nakamura, R. Arita, and M. Imada: J. Phys. Soc. Jpn. \textbf{79} (2010) 044705. 
\bibitem{aryasetiawan1}F. Aryasetiawan, M. Imada, A. Georges, G. Kotliar, S. Biermann, and A. I. Lichtenstein: Phys. Rev. B \textbf{70} (2004) 195104.

\bibitem{souza}I. Souza, N. Marzari, and D. Vanderbilt: \textit{ibid} \textbf{65} (2001) 035109.
\bibitem{marzari}N. Marzari and D. Vanderbilt: Phys. Rev. B \textbf{56} (1997) 12847.


\bibitem{miyake3}T. Miyake, F. Aryasetiawan, and M. Imada: Phys. Rev. B \textbf{80} (2009) 155134.
\bibitem{hohenberg}P. Hohenberg and W. Kohn: Phys. Rev. \textbf{136} (1964) B864.
\bibitem{korn}W. Kohn and L. S. Sham: Phys. Rev. \textbf{140} (1965) A1133 .
\bibitem{methfessel}M. Methfessel, M. van Schilfgaarde, and R. A. Casali; in Lecture Notes in Physics, edited by H. Dreysse (Springer-Verlag, Berlin, 2000), Vol. 535.
\bibitem{mullar} K. A. Muller and H. Burkard: Phys. Rev. B \textbf{19} (1979) 3593.
\bibitem{benthem} K. van Benthem, C. Elsasser, and R. H. French: J. Appl. Phys. \textbf{90} (2001) 6156. 
\bibitem{lim} S.-G. Lim, S. Kriventsov, T. N. Jackson, J. H. Haeni, D. G. Schlom, A. M. Balbashov, R. Uecker, P. Reiche, J. L. Freeouf, and G. Lucovsky: J. Appl. Phys. \textbf{91} (2002) 4500. 
\bibitem{okamoto} S. Okamoto, and A. J. Millis: Phys. Rev. B \textbf{70} (2004) 075101. 

\bibitem{hedin}L. Hedin: Phys. Rev. \textbf{139} (1965) A796.

\bibitem{nakamura3}K. Nakamura, Y. Yoshimoto, Y. Nohara, and M. Imada: J. Phys. Soc. Jpn. \textbf{79} (2010) 123708.

\end{thebibliography}
\end{document}